\documentclass[fleqn,10pt]{wlscirep}
\usepackage[utf8]{inputenc}
\usepackage[T1]{fontenc}
\usepackage{color,mhchem}

\usepackage{amsmath}

\usepackage{amsfonts}

\title{Tilted black-Si: $\sim 0.45$ form-birefringence from sub-wavelength needles}

\author[1,*,+]{Darius~Gailevi\v{c}ius}
\author[2,*,+]{Meguya~Ryu}
\author[2]{ Reo~Honda}
\author[3]{Stefan~Lundgaard}
\author[2]{Tai~Suzuki}
\author[3]{Jovan~Maksimovic}
\author[3]{Jingwen~Hu}
\author[3,4]{Denver~P.~Linklater}
\author[4]{Elena~P.~Ivanova}
\author[3]{Tomas~Katkus}
\author[3]{Vijayakumar~Anand}
\author[1]{Mangirdas~Malinauskas}
\author[5]{Yoshiaki~Nishijima}
\author[1]{Mangirdas~Malinauskas}
\author[3]{Vijayakumar~Anand}
\author[3]{Soon~Hock~Ng}
\author[6,7]{K\c{e}stutis~Stali\={u}nas}
\author[2,*]{Junko~Morikawa}
\author[3,5,8]{Saulius~Juodkazis}

	\affil[1]   {Laser Research Centre, Faculty of Physics, Vilnius University, Saul\.{e}tekio Ave. 10, LT-10223 Vilnius, Lithuania}
	\affil[2]   {CREST - JST and School of Materials and Chemical Technology, Tokyo Institute of Technology, 2-12-1, Ookayama, Meguro-ku, Tokyo 152-8550, Japan}
	\affil[3]   {Optical Sciences Centre and ARC Training Centre in Surface Engineering for Advanced Materials (SEAM), School of Science, Swinburne University of Technology, Hawthorn, VIC 3122, Australiav}
	\affil[4]   {School of Science, RMIT University, Melbourne, VIC 3000, Australia}
	\affil[5]   {Institute of Advanced Sciences, Yokohama National University, 79-5 Tokiwadai, Hodogaya-ku, Yokohama 240-8501, Japan}
	\affil[6]  {Dep. de F\'{i}sica, Universitat Polit\`{e}cnica de Catalunya (UPC), Colom 11, E-08222 Terrassa, Barcelona, Spain}
	\affil[7]  {Instituci\'{o} Catalana de Recerca i Estudis Avan\c{c}ats (ICREA), Passeig Llu\'{i}s Companys 23, E-08010, Barcelona, Spain}
	\affil[8]  {World Research Hub Initiative (WRHI), School of Materials and Chemical Technology, Tokyo
	Institute of Technology, 2-12-1, Ookayama, Meguro-ku, Tokyo 152-8550, Japan}

\affil[*]{Correspondence: (D. G.) darius.gailevicius@ff.vu.lt, (M. R.) ryu.m.ab@m.titech.ac.jp, (J.~M.)~morikawa.j.aa@m.titech.ac.jp.}

\affil[+]{D.G. and M.R. contributed equally to this work.}

\begin{abstract}
The self-organised conical needles produced by plasma etching of silicon (Si), known as black silicon (b-Si), create a form-birefringent surface texture when etching of Si orientated at angles of $\theta_i < 50 - 70^\circ$ (angle between the Si surface and vertical plasma E-field). The height of the needles in the form-birefringent region following 15~min etching was $d\sim 200$~nm and had a 100~$\mu$m width of the optical retardance/birefringence, characterised using polariscopy. The height of the b-Si needles corresponds closely to the skin-depth of Si $\sim\lambda/4$ for the visible spectral range. 
Reflection-type polariscope with a voltage-controlled liquid-crystal retarder is proposed to directly measure the retardance $\Delta n\times d/\lambda\approx 0.15$ of the region with tilted b-Si needles. The quantified form birefringence of $\Delta n = - 0.45$ over $\lambda = 400-700$~nm spectral window was obtained. 
Such high values of $\Delta n$ at visible wavelengths can only be observed in the most birefringence calcite or barium borate as well as in liquid crystals. The  replication of b-Si into Ni-shim with high fidelity was also demonstrated and can be used for imprinting of the b-Si nanopattern into other materials.
\end{abstract}

\begin{document}

\flushbottom
\maketitle
%
%
\thispagestyle{empty}





\section{Introduction}

Nanotextured silicon (Si) surfaces - in particular, black-Si (b-Si) - have a range of useful properties for widespread applications including MEMS, energy storage and biotechnology: it is anti-reflective over the visible-infrared (VIS-IR) spectral range~\cite{16aplp076104}, rendering it useful for the manufacture of solar cells~\cite{solar,16semsc221,16aplp076104} or other energy applications. Currently, there is a growing need for efficient photo-thermal catalysts capable of reducing  CO$_2$ into fuel
~\cite{Ozin,catal} and nanotextured Si surfaces are an attractive candidate due to their anti-reflective properties. They also exhibit enhanced electron transport at the nano-sharp (curvature radius $\sim 10$~nm) tips which act as contact points as well as at the locations where near-field polarisation is perpendicular to the interface. bSi has also been used as a highly efficient bactericidal topography towards pathogenic Gram-positive and Gram-negative bacteria~\cite{13nc2838}, and can rupture the membrane of red blood cells~\cite{14jmcb2819}. The intrinsically hydrophilic surface can be silanized and rendered hydrophobic to tune organic molecule-surface interactions for surface enhanced Raman spectroscopy (SERS)~\cite{13ap907}~\cite{15oe6763}. Recently, optical binding/trapping of DNA on b-Si was demonstrated by inducing nonlinear optical effects inside the trapped material using very high 1~MW/cm$^2$ laser powers~\cite{17sr12298}. The trapping was aided by the high thermal conductivity of b-Si, which in turn, is due to the crystalline nature of the substrate
and increased surface area due to nanostructuring~\cite{old}. Additionally, bSi has been reportedly used as a field emitter electrode~\cite{field}. Another evolving field of interest is the application of tilted nanostructures that feature extrinsic chirality~\cite{Verbiest1996, Belardini2015} as optical bio-sensors ~\cite{16sr31796}.

\begin{figure}[tb]
\centering
\includegraphics[width=16cm]{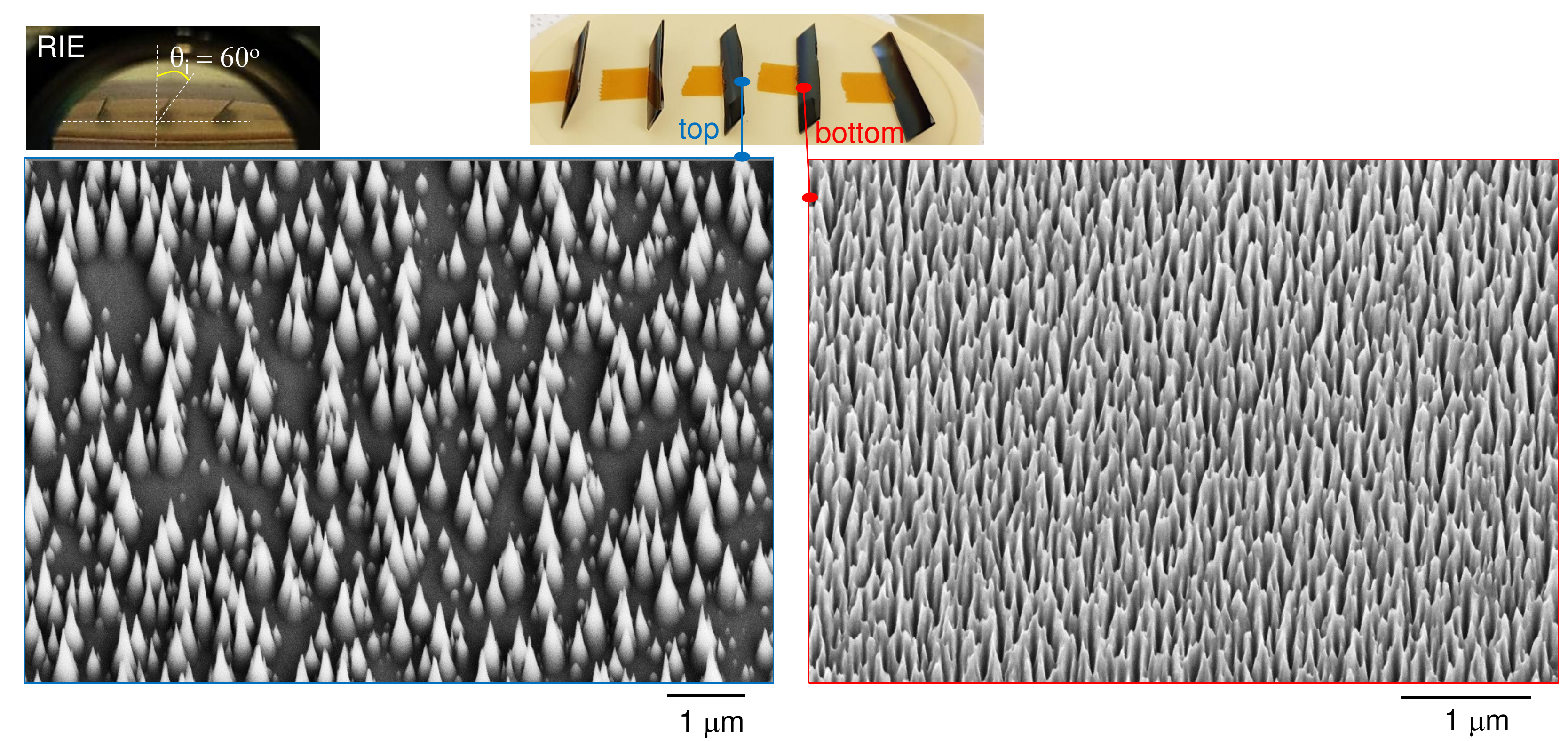}
\caption{Structural characterisation of b-Si with tilted conical structures by scanning electron microscopy (SEM). Plasma etching: electrical bias electrodes produced an E-field which directed at $\theta_i = 40-70^\circ$ angle to the surface of the Si wafer (left inset shows photo of samples inside an RIE chamber. The patterning of b-Si on the top regions were $\sim 100~\mu$m wide, made of separated cylindrical spikes (left) which evolved into a dense pattern of tilted cones (right); see detailed image in Fig.~\ref{f-grad}. Central inset shows etched samples polyimide-tape fixed to the alumina sample holder-plate used inside the RIE chamber.} \label{f-bla}
\end{figure}   

The refractive index $n$ of Si is 3.58 to 3.69 within the visible spectral window of $\lambda = 400$~nm to 800~nm, which is useful for the phase control of light in transmission/reflection modes or geometrical (in-plane) phase modulation using azimuthally patterned refractive index \cite{Levy2004, Lerman2008}. Waveplates~\cite{Hu2017} and optical vortex generators~\cite{Chong2015} (via geometrical phase) have been realised using micro-optical polarising elements patterned on thin Si films.  When acceptable by design, it is beneficial to use the reflection type of micro-optical elements (metasurface) due to doubling of the optical path. Hence, characterisation of such reflective optical elements with form-birefringent  nano-/micro-structured patterns of Si with complex refractive indices ($n+\mathrm{i}k$) is required. Reflection from pure dielectrics $n > 1$ (in air) will experience a $\pi$-shift in phase, while from metals ($n < 1$), phase is unchanged. A simple quantitative technique based on polariscopy is needed for the optical measurement of birefringence of sub-wavelength metasurface structures in the reflection mode. This was the major motivation for this study where we integrated the earlier developed multi-wavelength quantitative birefringence mapping method with the existing polarisation-microscope by the addition of a liquid crystal (LC) retarder and band-pass filters ($\sim\$1$k)~\cite{18sr17652}. 

Here, we demonstrate a large phase change ($\sim\pi/4$) of the reflected light from b-Si surface with tilted needles $d\approx 200$~nm in height. Plasma etching of Si at large angles creates the generation of nano-needles at an angle i.e. tilted b-Si. The properties of such surfaces were investigated in the visible-IR spectral ranges and showed low reflectivity $R = 2 - 4\%$. A birefringence corresponding to $\Delta n > 10^{-2}$ was estimated by cross-polarised imaging with $\lambda = 530$~nm plate retarder placed at a $45^\circ$-angle to introduce a color scale (the Michel-Levy birefringence color chart). Quantitative birefringence measurements showed the largest change of the birefringence corresponding to $\Delta n\approx 0.45$ over a $\sim 50~\mu$m wide region of b-Si where sparse tilted nano-needles formed a dense homogeneous pattern of tilted b-Si. High-fidelity replication of nano-textured b-Si with feature sizes 20 - 200~nm into a mechanically flexible Ni shim is also demonstrated for 3-inch Si wafers. Ni-shims can be used for transferring nano-textured surfaces into other materials including polymers.

\begin{figure}[tb]
\centering
\includegraphics[width=15cm]{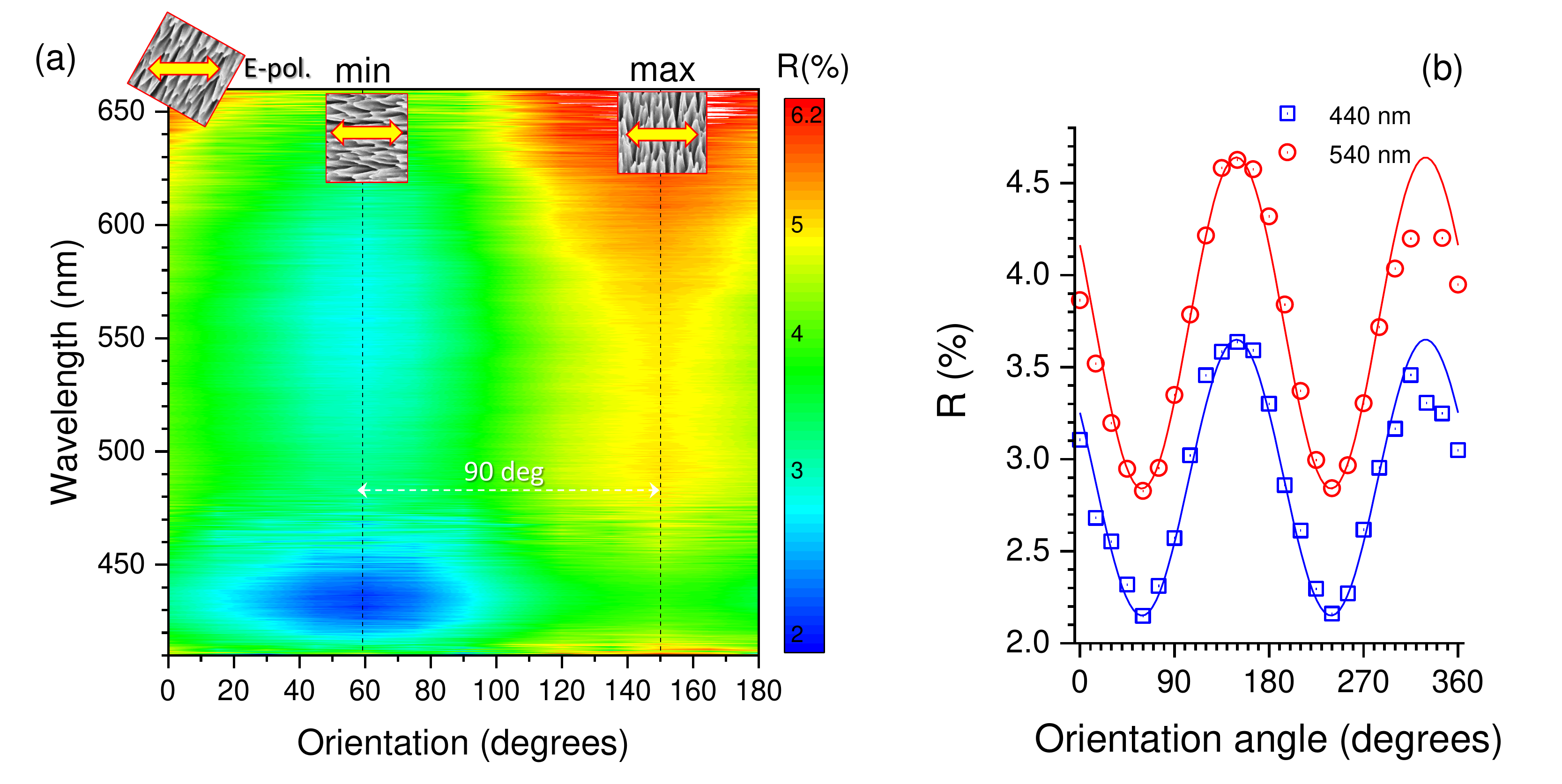}
\caption{(a) Reflection spectral map of b-Si region at different sample orientation angles $\theta$. Inset thumbnail images show respective orientation of incident E-field of light and the b-Si (see, Fig.~\ref{f-bla}). Note the $R$ is shown in logarithmic scale. (b) Reflectivity vs. orientation angle at the minimum of $R$ at $\lambda = 440$~nm and at the plateau region $\lambda = 540$~nm over $0-2\pi$ range; sample was rotated clockwise from original orientation $\theta_i = 0^\circ$. Lines are best fits by $\cos(2\theta)$ functions (is is equivalent to the $\cos^2\theta$ Malus dependence).  }\label{f-R}
\end{figure}

\begin{figure}[tb]
\centering
\includegraphics[width=12.cm]{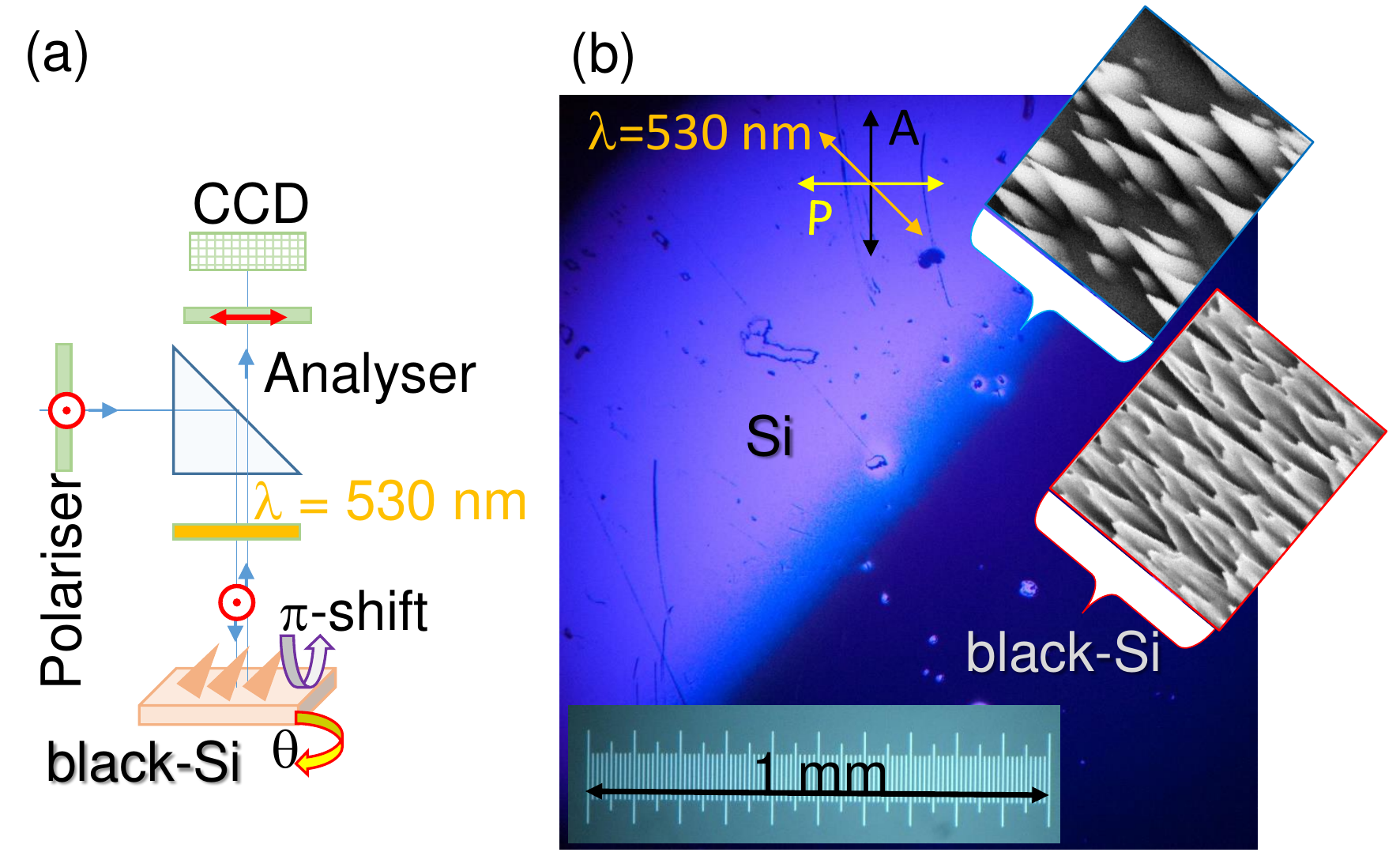}
\caption{(a) A cross-polarised microscopy schematic with a $\lambda = 530$~nm waveplate (for the color view) setup in reflection. The phase shift on reflection from larger refractive index surfaces of Si cause a $\pi$-shift. (b) Polarised image of the transitional region Si to b-Si (insets show schematically locations of the two distinct regions and their corresponding orientation). The objective lens used is with numerical aperture $NA = 0.3$ and the condenser illumination is by a LED lamp. Detailed SEM view of the gradient region is shown in Fig.~\ref{f-grad}. }\label{f-refl}
\end{figure}    

\begin{figure}[tb]
\centering
\includegraphics[width=16cm]{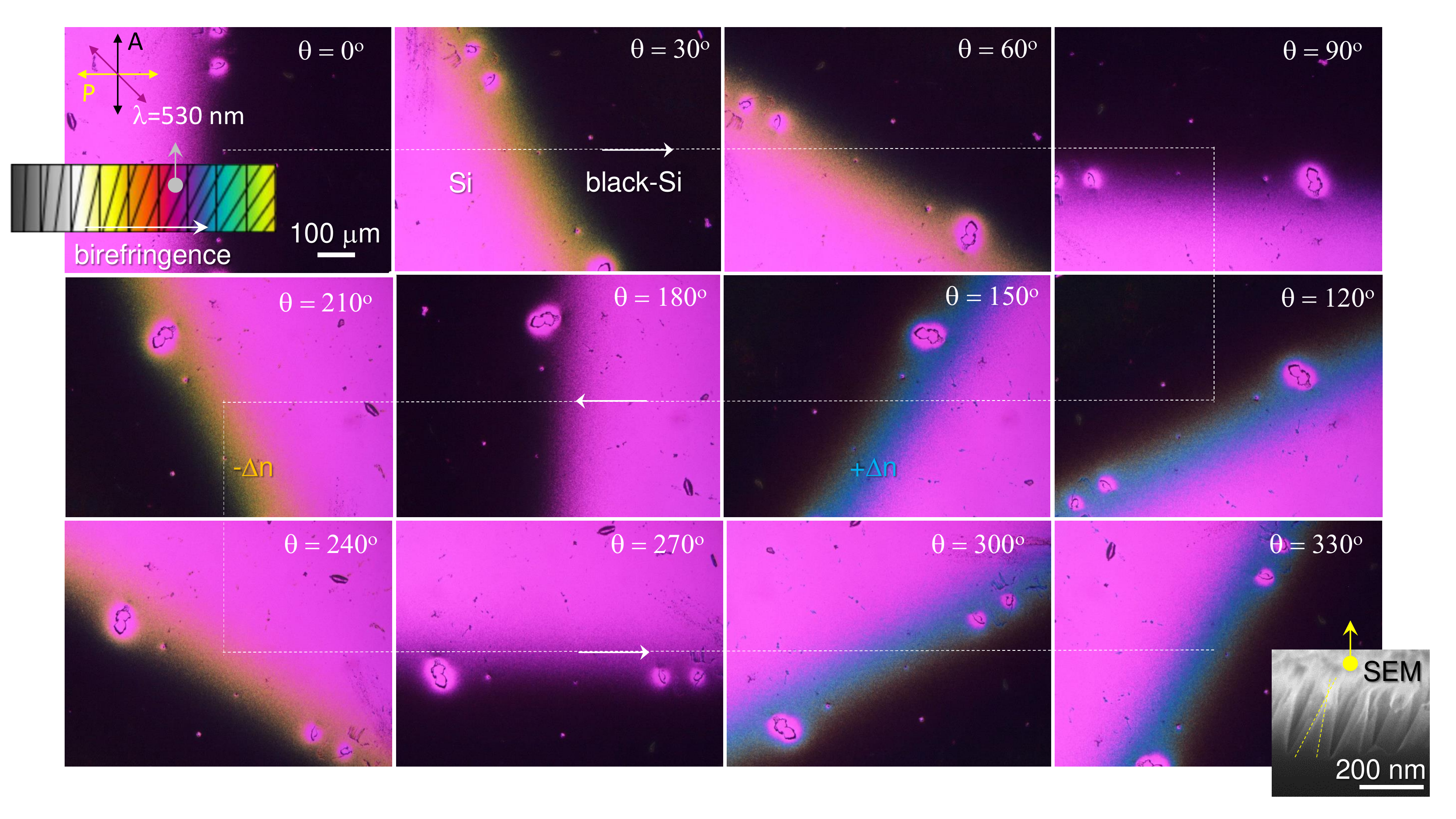}
\caption{Color-coded birefringence images with cross-polarised microscopy at different sample orientations (Fig.~\ref{f-refl}(a)) under halogen condensor illumination. The Michel-Levy birefringence color chart is shown in the inset of $\theta = 0^\circ$ image. Side-view SEM image of the tilted b-Si is shown at $\theta = 330^\circ$. The Nikon Optiphot-pol. microscope is equipped with an objective lens of numerical aperture $NA = 0.3$. See Fig.~\ref{f-cros} for the same region imaged without $\lambda = 530$~nm waveplate.}\label{f-shif}
\end{figure}

\begin{figure}[tb]
\centering
\includegraphics[width=13cm]{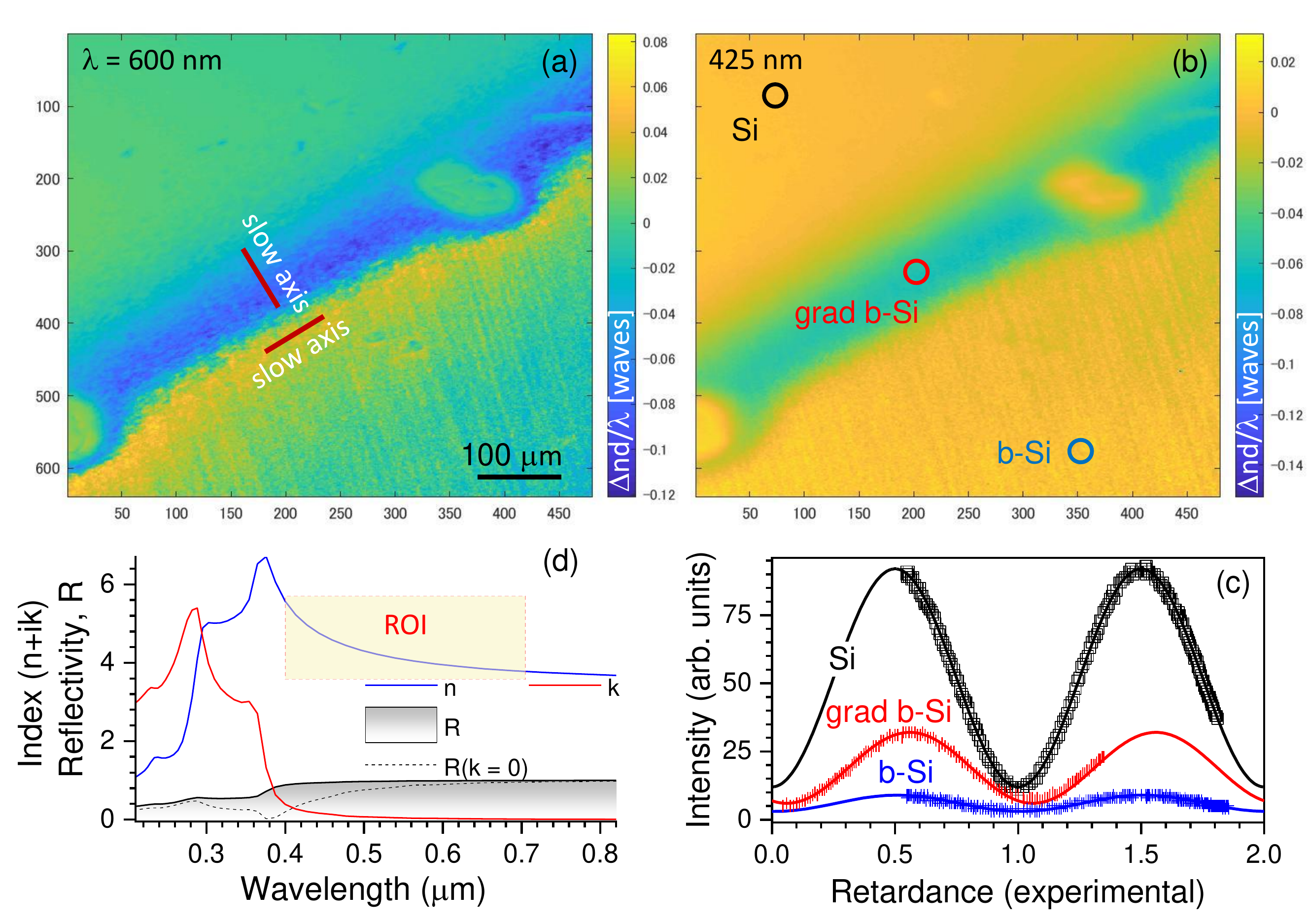}
\caption{Retardance maps for quantitative measurements. (a-b) The retardance at 600~nm and 425~nm from the region where flat Si region (top-left) is transferred into the gradient tilted b-Si with sparse nano-needles and dense b-Si (bottom-right). (c) Single pixel measurement of retardance by changing the voltage of the LC-retarder. The fit is made by $y=a\sin^2(\pi(x+\phi_{f})) + b$, where the phase difference $\phi_f$ is due to tilted b-Si needles; $\phi_f = 0^\circ$ as expected for flat Si. (d) Refractive index $(n + \mathrm{i}k)$ of Si and reflectivity $R$~\cite{refind}. The region of interest (ROI) marks the wavelength range where narrow band $\Delta\lambda = 20$~nm filters were used for quantitative measurements of retardance. }\label{f-reta}
\end{figure}    

\begin{figure}[tb]
\centering
\includegraphics[width=15cm]{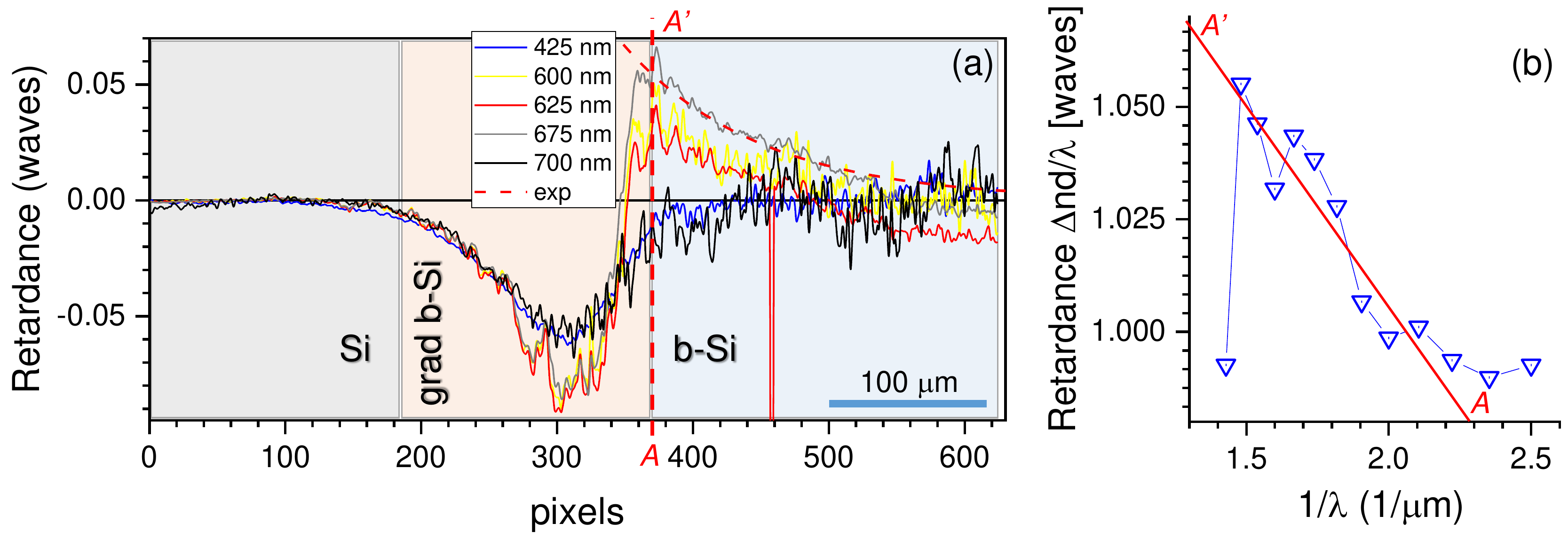}
\caption{(a) Cross-section normal of the gradient b-Si (Fig.~\ref{f-reta}(a,b) across the retardance change for different wavelengths. The exponential decay is added by dotted line for visual guidance. All cross sections are shown in Fig.~\ref{f-fobi}(b). (b) Dependence of $Retardance = f(1/\lambda)$ for $\lambda = 400 - 700$~nm at the \emph{A-A'} position. The slope corresponds to $\Delta n = -0.45$ for $d = 0.2~\mu$m along \emph{A-A'} cross section.  }\label{f-sect}
\end{figure}    

\section{Experiments}
\subsection{Fabrication of b-Si}

The b-Si was fabricated by plasma etching of clean 3" (inch) $\langle 100\rangle$ p-type (Boron doped) silicon wafers (University wafers, Ltd.) using a Samco RIE101iPH inductively coupled plasma (ICP) assisted reactive ion etching (RIE) tool. Silicon surfaces were initially cleaned with isopropanol and then dried under nitrogen gas flow to remove contaminants. The etchant gases were $\mathrm{SF}_6/\mathrm{O}_2$ with respective flow rates of 35/45~sccm. Process pressure was 1~Pa, ICP power of 150~W, bias power of 15~W~\cite{13nc2838} and etching time was 15-30~min. During the etch process, the spontaneous passivation mask formed on the surface as an etching by-product which provokes the formation of b-Si was removed by sonication in 10~wt\% sulphuric acid solution for 10~min~\cite{Jansen1995}.

Topographical characterisation was carried out by scanning electron microscopy (SEM). Optical polarisation microscopy was used to inspect dichroism in reflection by a fiber-coupled spectrometer (Ocean Optics). 

\subsection{Refection type polariscope}

A reflection-type polarisation microscope with a LC-retarder was assembled onto a Nikon Optiphot-pol microscope by adding a compartment for the LC-retarder (see Appendix for details; Fig.~\ref{f-niko}). A slow-axis of the LC-retarder was set perpendicular to the nano-needles of b-Si in the transition region from single needles to dense b-Si. The same measurement protocol used for the transmission-type polariscope~\cite{18sr17652} was applied for quantitative measurements of birefringence. The retardance $\Delta n\times d/\lambda$~[waves] was calculated for each pixel (VGA $480\times 640$) using a Matlab code. Since calculations for each pixel are independent, the parallel mode of calculation was implemented to shorten the data processing time by $\sim$20 times for the VGA data set from the original one for the sequential data processing protocol.

\subsection{Ni shim copy of b-Si}

The replication of b-Si into a Nickel (Ni) shim was performed with an initial magnetron sputtered deposition of a Ni seed layer on the b-Si. In order to ensure a continuous coating on the b-Si, a 5~nm of Cr layer was evaporated at a normal incidence and then a further 5~nm at $25^\circ$ from normal incidence as an adhesion layer and then ($\sim200$~nm) of Ni was sputtered at $0^\circ$. The b-Si samples were then mounted onto an acrylic sample holder using copper tape and attached electrically to a 200~mm diameter sample stage. Nickel plating was carried out in a nickel sulfamate based electrolyte (pH 4) in constant current mode at 9~A with voltage between 3-3.5~V. Plating was run for 6 hours, growing a $\sim 200~\mu$m thick Ni shim, similar to the thickness reported previously for replication of laser ablated surfaces~\cite{14pqe119}.

The b-Si was separated from the sample holder then cut out of the Ni sheet by mechanical large-radius bending of the entire shim. The shims were then immersed in 30\% KOH solution at $80^\circ$C for $\sim4$~hrs to remove the remaining Si. The resulting Ni-shim was a negative copy of the original 3-inch b-Si wafer.

\section{Results}

\subsection{Morphology and spectral characterisation of tilted b-Si}
Figure~\ref{f-bla} shows two typical regions of the b-Si surface texture when the Si is etched at a large angle $\theta_i$. The gradient transitional region from the pristine Si mirror-like surface into homogeneous b-Si texture was formed by well-separated conical nano-/micro-needles. Approximately 5-mm-wide Si strips were fixed on an alumina RIE carrier plate with polyimide tape at different tilt angles $\theta_i$ (Fig.~\ref{f-bla}). The b-Si recipe was used with etching times of 15-30~min. For normal incidence $\theta_i =\pi/2$, the 15~min etch time would form the typical vertical 250-280~nm tall silicon conical needles.  
 
Reflection of the b-Si region (right-side pattern in Fig.~\ref{f-bla}) was measured under microscope illumination at different sample orientation angles $\theta$ with linearly polarised illumination (Fig.~\ref{f-R}(a)). Low reflectivity $R< 5\%$ was observed over the entire spectrum with a minimum at $\lambda = 440$~nm. $R(\theta)$ at the minimum wavelength showed an angular dependence expected a wire-grid polariser, i.e., the  Mallus law as shown in Fig.~\ref{f-R}(b). The minimum $R$ was at the angle when b-Si cones were aligned with polarisation of light E-field (the strongest absorption in wire-grid). The maximum of $R$ was shifted by $\theta = \pi/2$ when the E-field was perpendicular to the surface projection direction of the cones (E-field perpendicular to the wire-grid). The Malus law $\cos^2\theta$ is observed in reflection (for the wire-grid it would correspond to the transmittance, $T$). The high refractive index of b-Si causes $\pi$-phase shift for reflection, however, does not introduce phase delay which would case ellipticity of polarisation as it was measured in a separate experiment (Appendix Fig.~\ref{f-diff}). Only minor differences for the reflection $R_\parallel$ and $R_\perp$ were observed (Fig.~\ref{f-diff}) indicating that there was no birefringence in the regions of sparse tilted b-Si cones etched at an angle relative to the normal of the samples's surfaces.   

\subsection{Birefringence of tilted b-Si}

To measure birefringence of the tilted b-Si, a polarisation microscope was assembled for the reflection mode (Fig.~\ref{f-refl}(a)). A reflection from high refractive index samples cause a $\pi$ phase shift but does not introduce anisotropy in reflection if the sample is not birefringent or dichroic.  By measuring reflection from the b-Si regions we show that $R_\parallel = R_\perp$ (Fig.~\ref{f-diff}). However, the transitional 100-$\mu$m-wide stripe between Si and b-Si had a gradient pattern of separated conical pyramids and is expected to be birefringent. By introducing a $\lambda=530$~nm waveplate at $\pi/4$ angle to the crossed polariser-P and analyser-A, the birefringence value is transferred into a color pallet (the Michel-Levy chart), which is more intuitive as compared to the black-white image observed in crossed P-A setup in transmission. However, Fig.~\ref{f-refl}(b) does not show expected differences of the colors at different orientations (one orientation is shown in (b)) due to monochromatic nature of LED lamp used for the illumination. With halogen white broadband spectrum illumination, a distinct color change with orientation was observed (Fig.~\ref{f-shif}). Color changes indicated presence of birefringence even larger than that observed in the laser-inscribed sapphire where the length of stressed volume was $d\approx 40~\mu$m and $\Delta n\propto 10^{-4}$~\cite{19n1414}. Here, $d = 0.2~\mu$m and proportionally higher birefringence is expected for the same retardance $\Delta n\times d$ (the same color on the Michel-Levy chart). An extraordinary high birefringence corresponding to $\sim\lambda/5$ retardance can be explained by the effective refractive index of the reflecting region. The skin depth $\sim\lambda/4\approx 200$~nm contributes to the reflection. This form-birefringent region which by definition has negative birefringence $\Delta n < 0$ acts effectively as medium with $n < 1$. This changes the phase change upon reflection from air ($n = 1$), hence the retardance is changed. 
Figure~\ref{f-cros} clearly shows that only gradient index region of tilted b-Si is birefringent. The Si and dense b-Si sides are black in the cross-polarised image. 

\subsection{Quantitative measurement of birefringence in reflection}

To quantify the retardance (hence, birefringence), a liquid crystal (LC) retarder was introduced in front of the CCD (Fig.~\ref{f-refl}(a); see also Fig.~\ref{f-niko}) and the same measurement protocol was used as developed for the transmission type of birefringence measurements~\cite{18sr17652}. This upgraded setup was necessary, since using a Berek compensator it is not possible to determine the birefringence. This is partially due to a strong reflection from unstructured regions of mirror-like Si.

An orientation of the sample was set with projection of tilted needles perpendicular to the slow-axis of LC retarder. The voltage of LC-retarder was changed in small steps for $N = 700$ frames taken by CCD; an initial calibration was made to prevent saturation of the CCD and to use entire dynamic range of the 8-bit camera. Figures~\ref{f-reta}(a,b) show retardance $\Delta n\times d/\lambda$~[waves] maps calculated for every pixel of the image. An intensity of each CCD pixel measured for $N=700$ voltage points of LC-retarder are shown in Fig.~\ref{f-reta}(c). The intensity was fit with the $y=a\sin^2(\pi(x+\phi_{f})) + b$ function (lines in  Fig.~\ref{f-reta}(c)). There is an apparent phase shift $\phi_f$ between retardance moving from the flat Si region towards increasingly denser concentrations of tilted b-Si nano-needles. This shift depends mostly on the birefringence $\Delta n$ since dichroism $\Delta\kappa$ is considerably smaller due to negligible $\kappa\,\to\,0$ (Fig.~\ref{f-reta}(d)). It is noteworthy that Si has increasing $\kappa$ at shorter wavelengths and the reflectivity has a contribution due to $n$ as well as $\kappa$ (see specific contributions in Fig.~\ref{f-reta}(d)). The cross-polarised measurements used here are sensitive to $\Delta n$  and dichroism $\Delta\kappa$ when extinction $\kappa$ increases. Separation of $\Delta n$ and $\Delta\kappa$ related contributions to the angular  anisotropy of the transmitted signal can be separated by sample rotation if required as we demonstrated recently~\cite{19nh1443}. In this study, this was not carried out since birefringence change dominated the transmission in cross-polarised measurements. In the cross polarised imaging with the LC-retarder, dichroism is reducing $E_{x,y}$ fields along slow (x-axis) and fast (y-axis) axes without introducing phase between the field components as it is the case for the birefringence.      

Cross sections normal to the interface between sparse tilted Si needles and dense b-Si were extracted from the retardance maps (Fig.~\ref{f-reta}(a,b)) at several wavelengths as a single pixel cross sections. The expected retardance $\Delta n\times d/\lambda = 0$ was observed on flat Si and as an asymptotic value on the b-Si. A strong swing of retardance was observed on the gradient region of the tilted sparse nano-needles (Fig.~\ref{f-sect}). A typical tendency was that a spatial evolution of retardance throughout the gradient region followed the same slope for all the wavelengths (Fig.~\ref{f-sect}(a)), then after a minimum, it crossed into positive retardance values and was wavelength dependent. Polariscopy is not sensitive to the absolute value of the phase (upon reflection in our experiment), hence, an advance of light wave (a negative retardance) is due to reflectivity by tilted nano-needles in the gradient region which are in front of the Si flat background. Since the sample is not rotated during measurements and only the LC-retarder is imparting a changing phase during measurements, the tilted nano-needles are absorbing and re-emitting all the wavelengths in the absorption region of Si with the same phase in the gradient region. As a result, a similar retardance was observed for all the wavelengths $\lambda = 400 - 700$~nm and were with negative values corresponding to $\sim 5\%$ of the wavelength.    

To analyse the region of strong retardance dependence and its cross over into positive values, we applied a refractive index analysis by plotting the retardance $\propto f(1/\lambda)$ at the location close to the maximum values (along \emph{A-A'} cross section Fig.~\ref{f-sect}(a)).  The results in Fig.~\ref{f-sect}(b) show negative slope which corresponds to the $\Delta n = -0.45$ for the $d = 200$~nm height of the b-Si. This is an extraordinarily high value and the negative sign is consistent with the form-birefringence of b-Si, which is negative by definition~\cite{19n1414}. This birefringence caused a phase change of $\sim 6-7\%$ of the wavelength across the full visible spectrum. In space of a few tens-of-micrometers, the retardance upon reflectance changed between 10-15\% of the wavelength over a wide visible spectral window 400-700~nm. Since many optical sensors are based on detection of refractive index changes (or retardance), the gradient field of tilted b-Si can be useful for recognition of the attached analyte that changes the corresponding slope of retardance vs wavelength as shown in Fig.~\ref{f-sect}(b).

\begin{figure}[tb]
\centering
\includegraphics[width=13cm]{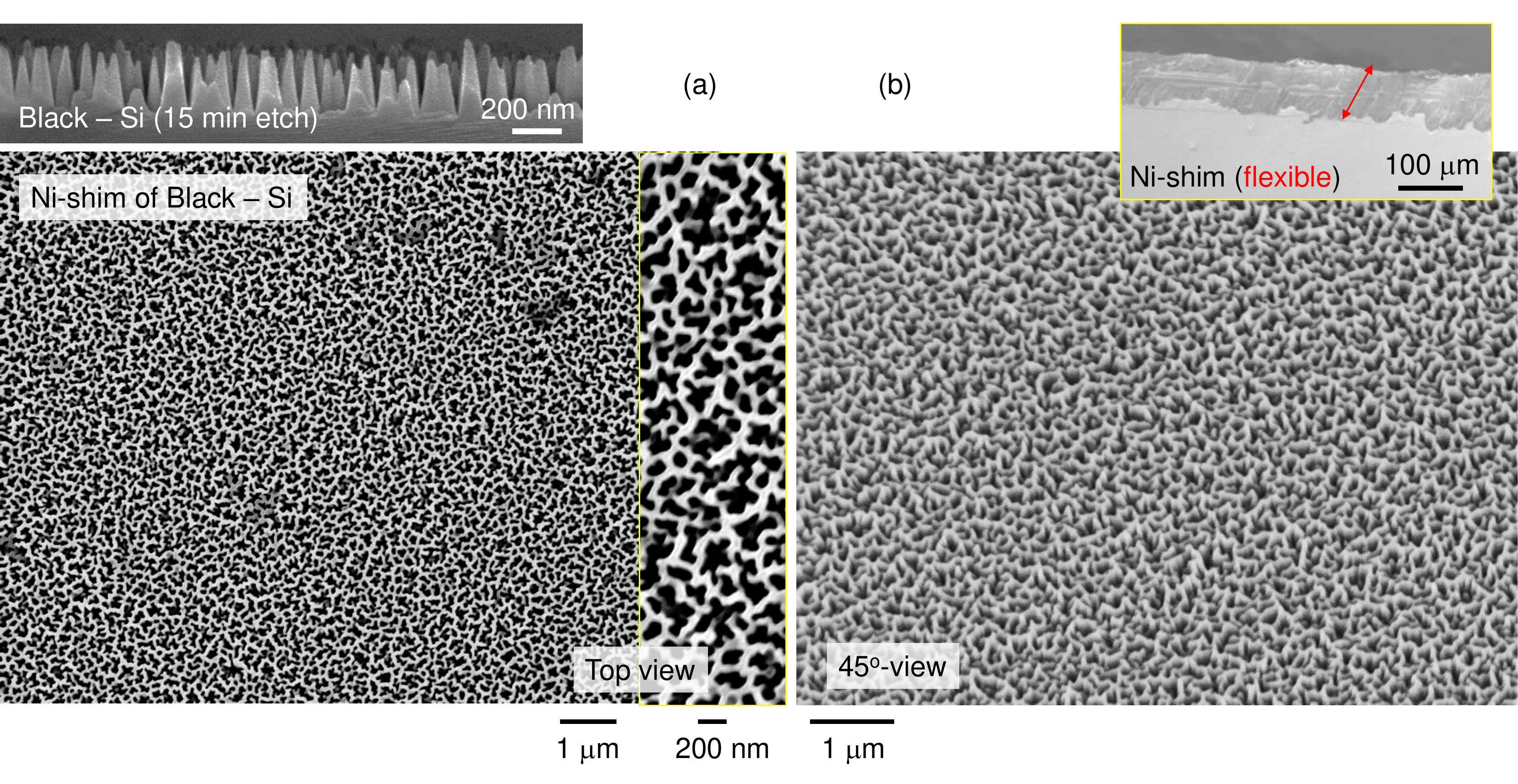}
\caption{SEM top- and side-view images of Ni shim made from b-Si with vertical cones etched by the same recipe as the tilted b-Si for 15~min.     }\label{f-Ni}
\end{figure}    

\subsection{Replication of nanotextured b-Si}
 
A nanotextured surface of b-Si has number of applications and development of nano-texture transfer into other materials, e.g., to create biocidal plastics with needle separation $\sim 100$~nm or fabrication of nano-textured (black) photo-catalytic surfaces for water splitting~\cite{green} are required. Here, the first step is accomplished by transfering b-Si nanotexture into a Ni shim copy. This was earlier demonstrated for laser ablated surfaces with replication of features down to 20~nm~\cite{14pqe119}. Since Si was chemically dissolved from the Ni-shim, it was possible to use this method to copy b-Si with vertical and tilted cone shaped patterns. Figure~\ref{f-Ni} shows top and slanted-view SEM images of the b-Si made by 15~min plasma etch of 400-$\mu$m-thick 3-inch $\langle 100\rangle$ Si wafer. Electrochemical deposition of a 200-$\mu$m-thick Ni shim took 6~hours. Ni shims are a good choice for nano-imprint and roll-to-roll replication of nano-patterns due to their ductility and chemical stability. 

\section{Discussion}

The demonstrated retardance change was observed for a reflected beam and is due to a delicate change of the refractive index in a transitional manner. The observation were made with a new polariscopy setup with an integrated LC-retarder. The gradient pattern of the transitional region from pristine Si to b-Si is favorable to observe the location where the change of effective refractive index is taking place and to determine its value. Also, Si solar cells which usually have an antireflective coating might benefit from engineering surface pattern which controls the phase of the reflected beam and facilitates stronger absorption at the pre-surface regions. 

Interestingly, the fabricated tilted nano-needle patterns are chiral and can have applications in Raman scattering/spectroscopy. Figure~\ref{f-fdtd} shows strong localisation of light at the very tips of the nano-needles due to the oblique angle with the incident E-field and high n of Si. In contrast, for needles normal to the surface there is no light enhancement observed. The field enhancement compared with vertical (not shown for brevity) needles in terms of intensity $E^2$ is $\approx 14^\times$. We could hypothesise here that such strong localisation of light on the tips can be useful for bactericidal and biocidal (also antiviral function of nanotextured surfaces of b-Si~\cite{13nc2838}.  

\section{Conclusions and outlook}

It is demonstrated that the $d = 200$~nm thick nanotextured Si surface is effectively a birefringent reflecting surface with a $\Delta n\approx 0.4$ at the visible spectral range. Effectively, a $\lambda/4$ ($\pi/2$ phase) change can be achieved with the nanotextured Si having a tilted feature pattern. Such pattern is produced using a simple plasma etching at an oblique angle approach. This negates the need for more complex multi-step lithography methods used for fabricating metamaterials for phase and amplitude control. The suggested technique could be appealing for a number of applications in sensing and harvesting of solar energy. Future efforts are needed to establish etching parameter (chemistry, etching angle and time) range for birefringence control. For the naturally birefringent materials, e.g., calcite, quartz, the proposed method of tilted etching can enhance that property creating meta-birefringent materials. 

\section*{Funding}
This work was supported by JST CREST Grant Number JPMJCR19I3, Japan,  the ARC Discovery DP190103284, and NATO No.SPS-985048 grants. 

\section*{Acknowledgments}
Support of operational costs of Nanotechnology facility by Swinburne Univ. Technol. (2016-19) is acknowledged. We are grateful to Dr. Ignas R\.{e}klaitis for discussions on plasma etching at angle.

\section*{Disclosures}
The authors declare no conflict of interest.

\renewcommand{\thefigure}{A\arabic{figure}}
\section*{Appendix}
\setcounter{figure}{0}  
\textbf{Gradient nano-texture.}  Figure~\ref{f-grad} shows detailed SEM image through the transition region from flat Si to dense b-Si. The gradient field of tilted nano-needles was observed in the location of the Si sample which was the farthest from the bottom plane of the sample carrier plate (more into the plasma chamber between the electrodes). An approximate extension of that region was 100~$\mu$m. 

\textbf{Polarisation dependence of IR reflectivity.}  Figure~\ref{f-diff}(a) shows reflectivity at two perpendicular orientations from the gradient b-Si region at visible spectral range. The difference due to the dichroism was within an uncertainty range recognisable only as a centre shift from the average (see blue vs red dot distribution). For IR (Figure~\ref{f-diff}(b)), there was no recognisable difference for reflections at different orientations when measurements were carried out at gradient and dense b-Si regions (only two polarisations are shown in (b) for better presentation). The difference in spectral behaviour of $R$ between regions was well distinct by change of $R$ from 15\% to 0\% at the longer wavelengths. b-Si has lower reflectivity due to the lower effective refractive index.       

\textbf{Visualisation and quantification of birefringence.}  Figure~\ref{f-cros} represents the cross-polarised image of the gradient b-Si region at different orientation angles $\theta$ and represents direct proof of the formation of strong birefringence from only 200-nm-tall tilted needles. The same sample region at the same magnification is shown in Fig.~\ref{f-shif} with $\lambda = 530$~nm waveplate to introduce colour pallet to $-|\Delta n|$ and $+|\Delta n|$.  

The modified polarisation microscopy setup with addition of LC-retarder is shown in Fig.~\ref{f-niko}. This setup enables measurements to be carried out at different orientation of the slow-axis of the LC-retarder when the sample has a changing orientation pattern. In our study, the orientation of nano-needles was fixed by plasma etching geometry: sample loading angle and plasma etch directionality and measurements, which were carried out at one fixed rotation angle of the sample. 

\textbf{Chirality and light field enhancement.}  Tilted needle metasurface patterns have  extrinsic/pseudo-chirality~\cite{16sr31796} which can be exploited by setting light-matter interaction which lacks mirror symmetry, hence, are chiral. Finite difference time domain (FDTD) calculations were carried out for two opposite tilt angles of Si cones at different 
created due to combination of high refractive index of Si and slanted incidence of light. Lower field enhancement ($E/E_0 = $2.9 vs 3.7) was obtained for the same pattern as shown in (Fig.~\ref{f-fdtd}) only for cones normal to the Si surface (not shown for brevity).   

\textbf{Anisotropy and dichroism.} Two effective medium approaches are compared next to model birefringence with results shown in Fig.~\ref{f-fobi}. In any type of material, a modification of adjacent regions with resolution smaller than the wavelength can be used to change the effective refractive index which for extraordinary $n_e$ (along the optical axis) and ordinary $n_o$ indices depends on the volume fraction $f=w/\Lambda$ in a grating structure:
\begin{equation}\label{En}
n_e = \sqrt{\frac{n_1^2n_2^2}{(1-f)n_2^2 + fn_1^2}};~~~n_o=\sqrt{(1-f)n_1^2 + fn_2^2},
\end{equation}
where $n_1$ and $n_2$ are the indices of the air and tilted b-Si needles, respectively, $w$ is width of one region (air) and $\Lambda -w$ is the part occupied by Si needles ($\Lambda$ is the period of the structure). Figure~\ref{f-fobi}(b) shows the form-birefringence calculated for a grating pattern of alternating air-Si at different volume fractions of Si. A large negative $\Delta n$ can be obtained. However, this approach makes an overestimate in the case of tilted b-Si since the structure is not a grating. A more precise estimate is discussed next. 

To estimate the anisotropy - a dependence of the refraction index and absorption on polarisation direction - for the tilted b-Si structure, we adapt the Bruggeman Efficient Media Theory (EMT) theory~\cite{Bruggeman1935,choy2015effective} for anisotropic composites. Hinted by the photos in Fig.~2, we consider the nanostructured b-Si surface as the anisotropic inclusion in air. Anisotropy factor of the structure was estimated from the autocoerellation function: $K(\vec{r})=\int \rho (\vec{r}_1-\vec{r}) \rho (\vec{r}_1) d\vec{r}_1 / \int \rho (\vec{r}_1)^2 d\vec{r}_1$. Figure~\ref{f-fobi}(a) shows the density plot of the autocorrelation function calculated (see Eqn.~3 in ref.~\cite{Robertson2012}) for the case shown in Fig.~2(a). The estimated anisotropy $k  = a/b$ of the structure is $k \approx 3$. Bruggeman EMT provides the implicit expression of the complex refraction index of composites consisting of media with susceptibilities $\epsilon_1$ and $\epsilon_2$ with filling factors, which are $(1-f)\approx38\%$ and $f \approx 62\%$ (Si), respectively, for our case~\cite{Bruggeman1935}:
\begin{equation}\label{Brugeman}
\frac{(1-f)(\epsilon_1-\epsilon)}{\epsilon_1+k\epsilon}+\frac{f(\epsilon_2-\epsilon)}{\epsilon_2+k\epsilon} = 0,
\end{equation}
here $k$ is the above estimated anisotropy factor, $\epsilon_1 = 1$ stands for air, and $\epsilon_2 = 15.7+0.16\mathrm{i}$ for silicon. The Bruggeman EMT provides the complex refraction indices for polarisations along the anisotropy axes (considering given $k$ ) and perpendicularly to axis (considering $1/k$ instead of $k$). For isotropic composites in 2D case $k = 1$, for maximally anisotropic structure (the stripes), $k = 0$ and $k \xrightarrow{} \infty$ for both polarisations, respectively. In our case, the anisotropy is partial and estimates in this particular case are $k = 3$ and $k = 1/3$.
The complex refraction indices $\tilde{n}=n+\mathrm{i}\kappa=\sqrt{\epsilon}$ are: $\tilde{n} = 2.92 + i\times 0.014$ for $k = 3$ (ordinary beam) and $\tilde{n} = 1.89 + i\times 2.89\times 10^{-3}$ for $k = 1/3$ (extraordinary beam), when $f = 0.62$ (Si fraction). Hence, the birefringence $|\Delta n|\equiv |n_e - n_o|= 1.03$  and dichroism $\Delta\kappa = 0.011$ is estimated for the $k = 3$ vs. 1/3 anisotropy (ordinary vs. extraordinary polarisation components). The dichroism has a smaller effect compared to the birefringence. The estimate of birefringence is approximately twice larger as the experimentally determined value, however, is smaller as compared with the estimate for the grating Eqn.~\ref{En} shown in Fig.~\ref{f-fobi}(b). 
For comparison, in the isotropic $k = 1$ case (Eqn.~\ref{Brugeman}) one finds $\tilde{n} = 2.47+i\times 9.2\times 10^{-3}$, while for the maximum anisotropy $k = 0$:  $\tilde{n} = 1.544+i\times 7.41\times 10^{-4}$. Bruggeman's EMT provides a better model for the tilted b-Si. 


\begin{figure}[tb]
\centering
\includegraphics[width=12cm]{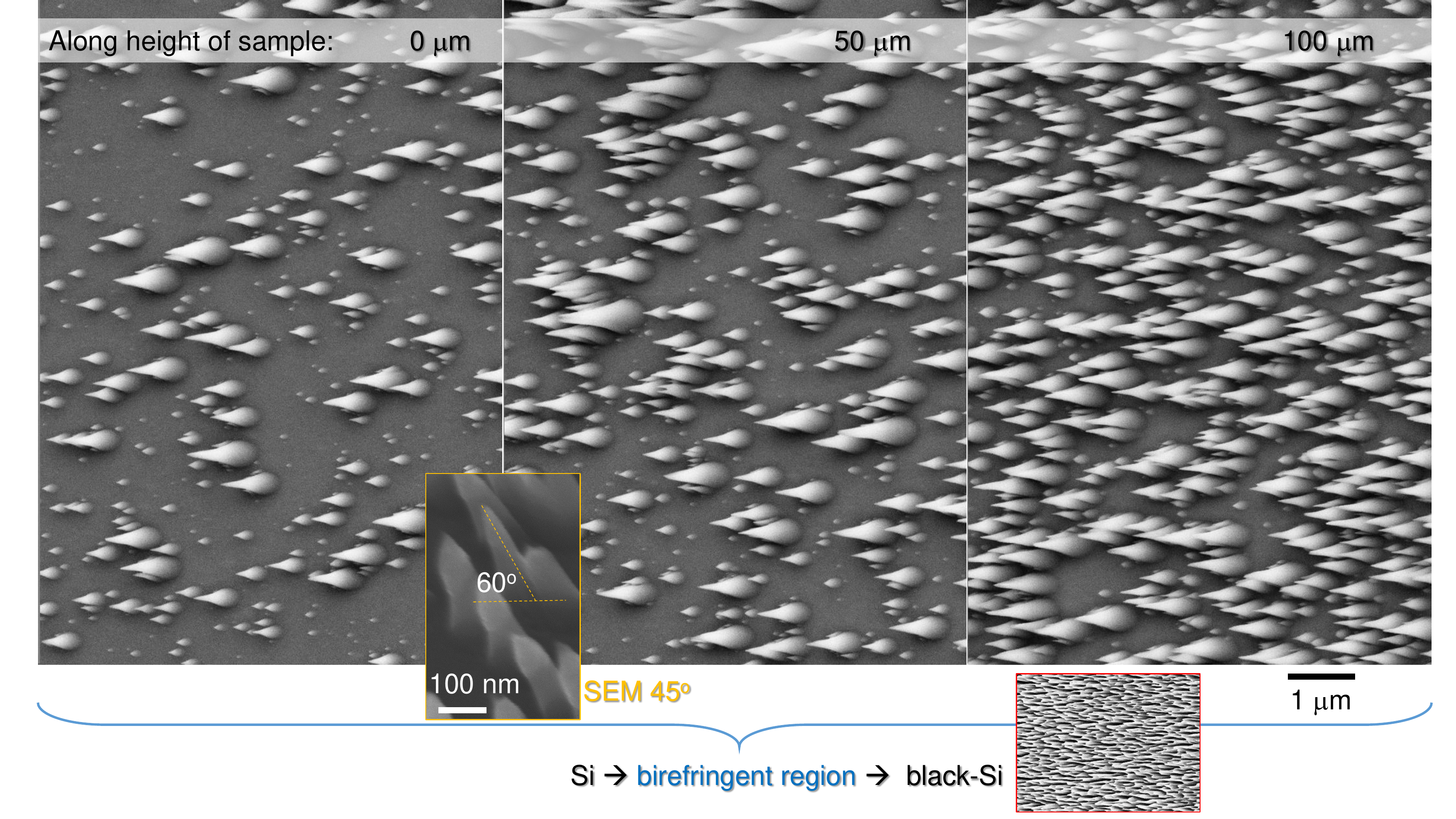}
\caption{SEM images of the gradient region of tilted b-Si. Region shown on the left-side was deeper into the RIE etching chamber (more far away from the sample holder). }\label{f-grad}
\end{figure}    
\begin{figure}[tb]
\centering
\includegraphics[width=14.5cm]{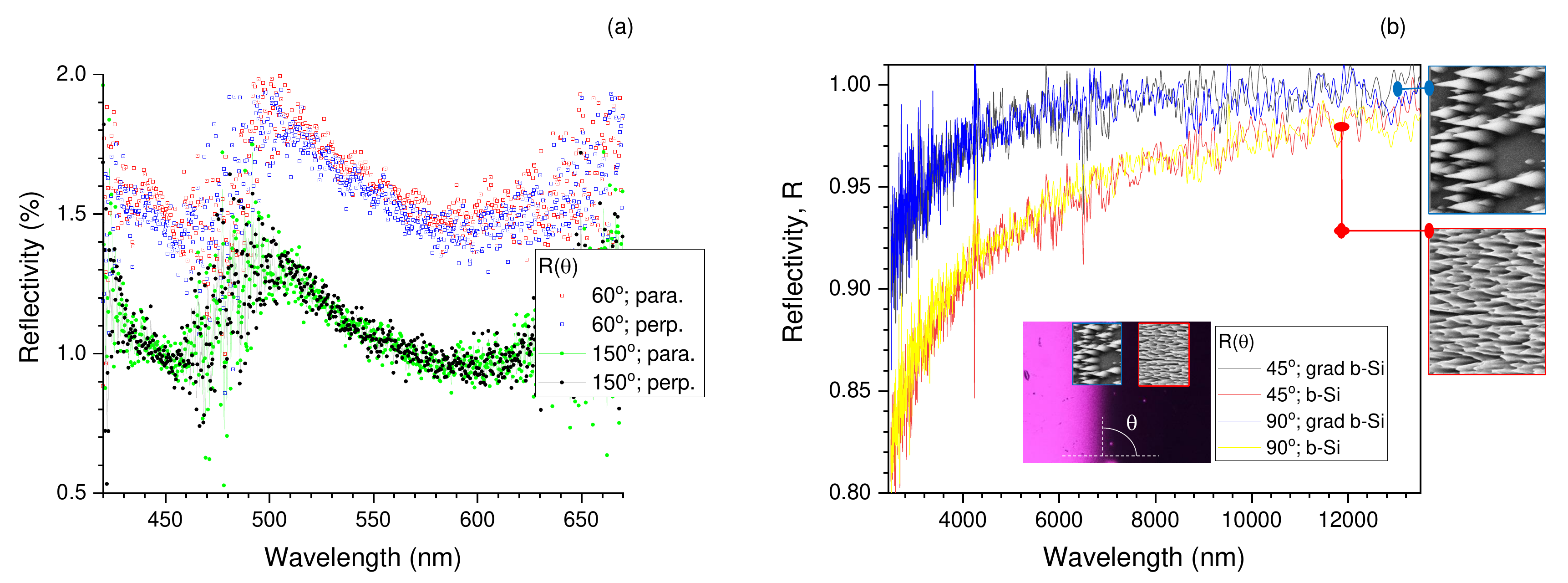}
\caption{Reflectivity in visible-IR. (a) Reflectivity from the tilted b-Si region measured in two perpendicular orientations parallel and perpendicular to the E-field of illuminated light, $R_{\parallel,\perp}$, respectively, at the sample orientation $\theta=60^\circ$ ($R$ minimum) and $\theta=150^\circ$ ($R$ maximum) (see Fig.~\ref{f-R}(b)). (b) Reflectivity measured by FTIR spectrometer at different incident polarisations ($\theta = 0^\circ$ correspond to the horizontal pol.;  see the inset for orientation definition.) from the gradient and dense b-Si. }\label{f-diff}
\end{figure}  

\begin{figure}[tb]
\centering
\includegraphics[width=13.5cm]{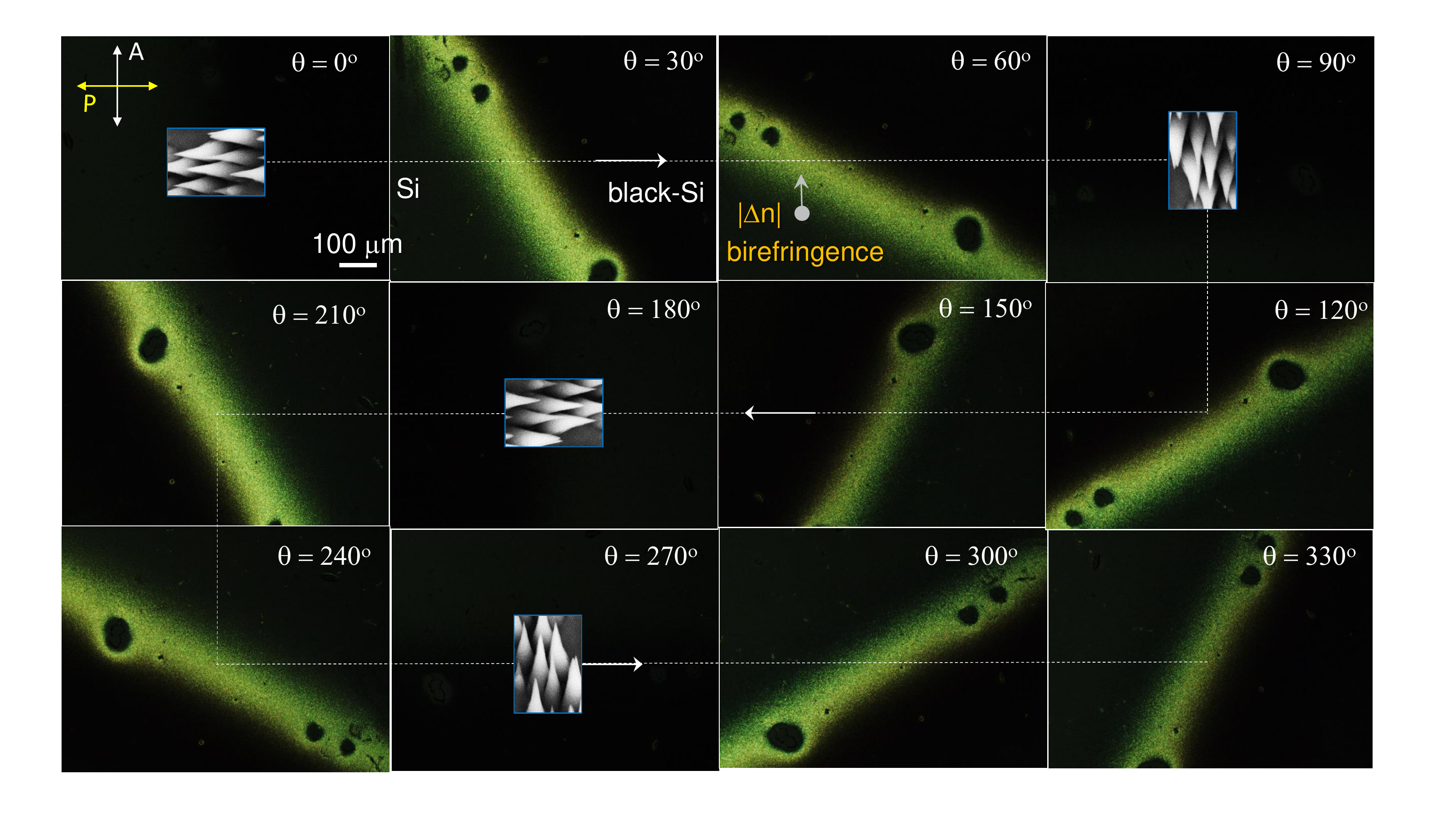}
\caption{Cross-polarised images of the same tilted black Si region at different $\theta$  angles as shown in Fig.~\ref{f-shif} (without $\lambda = 530$~nm waveplate); illumination is by halogen lamp. The birefringence is recognisable only in the gradient region. Insets show SEM-thumbnail images to highlight the orientation of tilted b-Si needles; see the crossed and aligned orientations with analyser or polariser (at the dark images). Nikon Optiphot-pol. microscope with objective lens of numerical aperture $NA = 0.3$. }\label{f-cros}
\end{figure}    

\begin{figure}[tb]
\centering
\includegraphics[width=7cm]{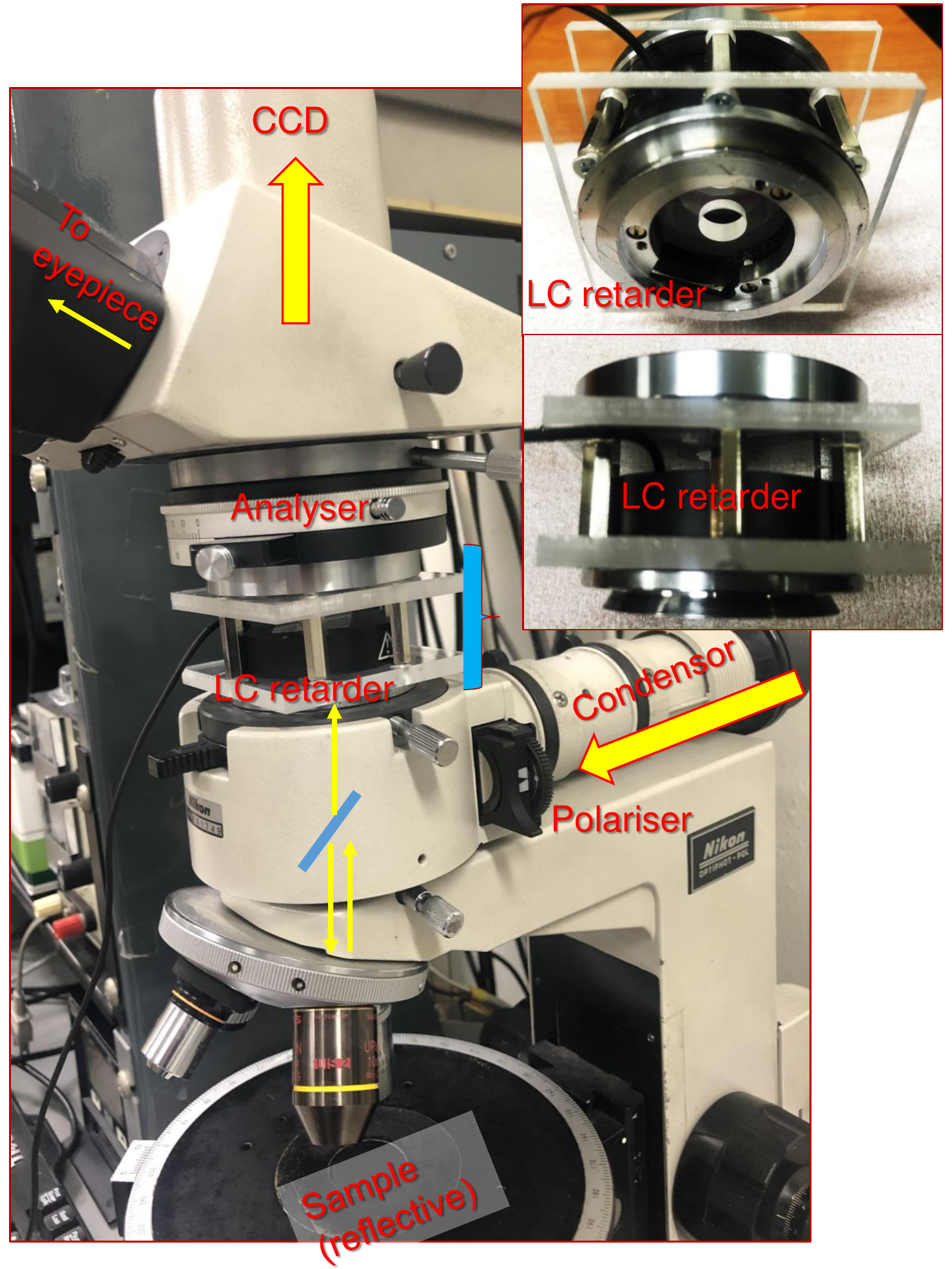}
\caption{Polariscopy setup based on the Nikon Optiphot - pol. microscope modified with a liquid crystal (LC) retarder placed in front of CCD camera. Insets shows the LC housing with standard Nikon fixation mounts: side and slanted-view photos. The same mount can be used for the usual transmission mode of polariscope~\cite{18sr17652}. }\label{f-niko}
\end{figure}   
\begin{figure}[tb]
\centering
\includegraphics[width=12cm]{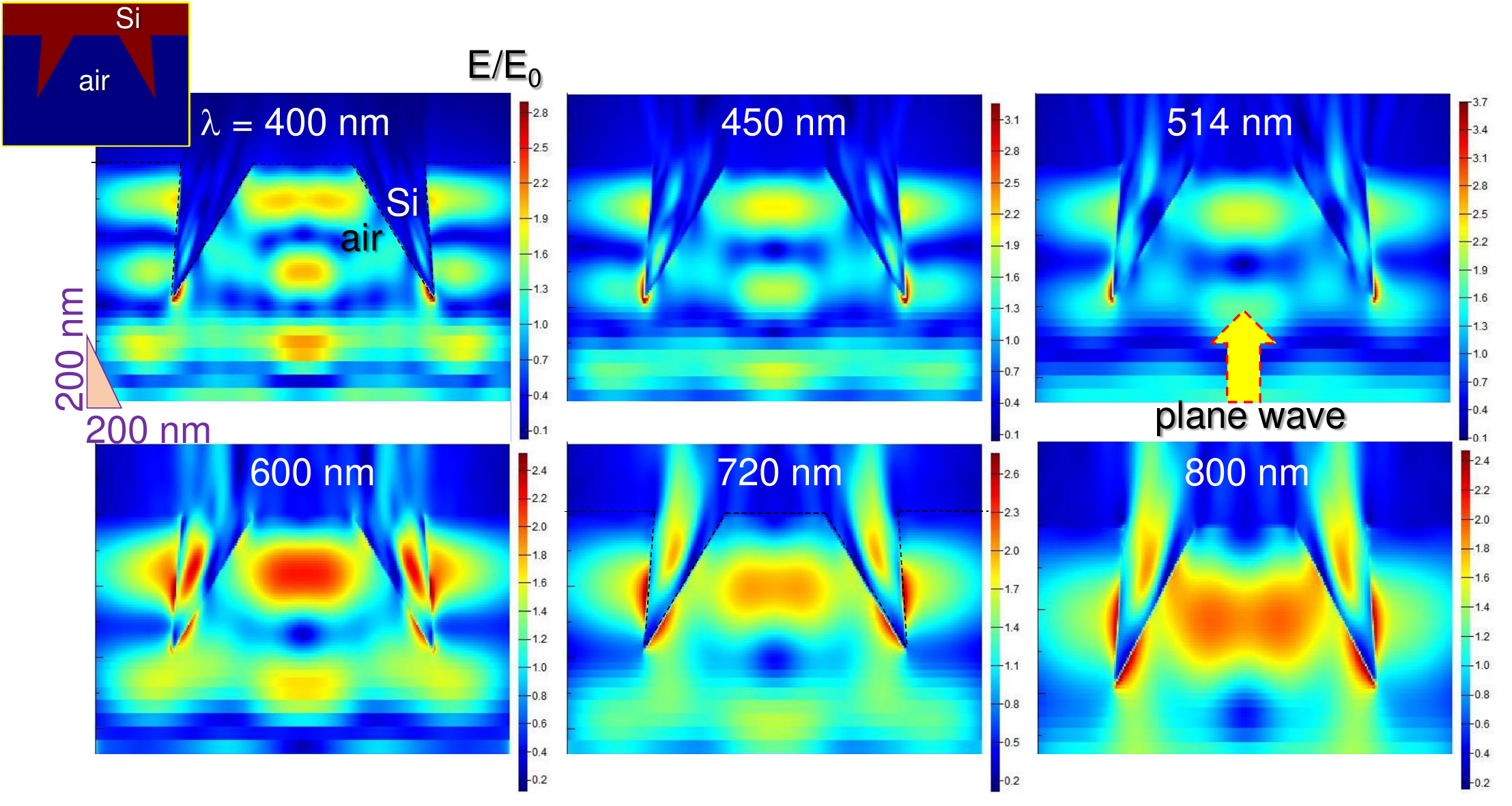}
\caption{Light enhancement on Si nano-needles at 30$^\circ$-tilted cones of two opposite inclination at different wavelengths; incident field $E_0 =1$. Noteworthy, this pattern is chiral. Finite differences time domain (FDTD) calculations were carried out with Lumerical Solutions program; the refractive index $(n(\lambda)+\mathrm{i}\kappa(\lambda))$ of Si is taken from Lumerical database.}\label{f-fdtd}
\end{figure}   
\begin{figure}[tb]
\centering
\includegraphics[width=15cm]{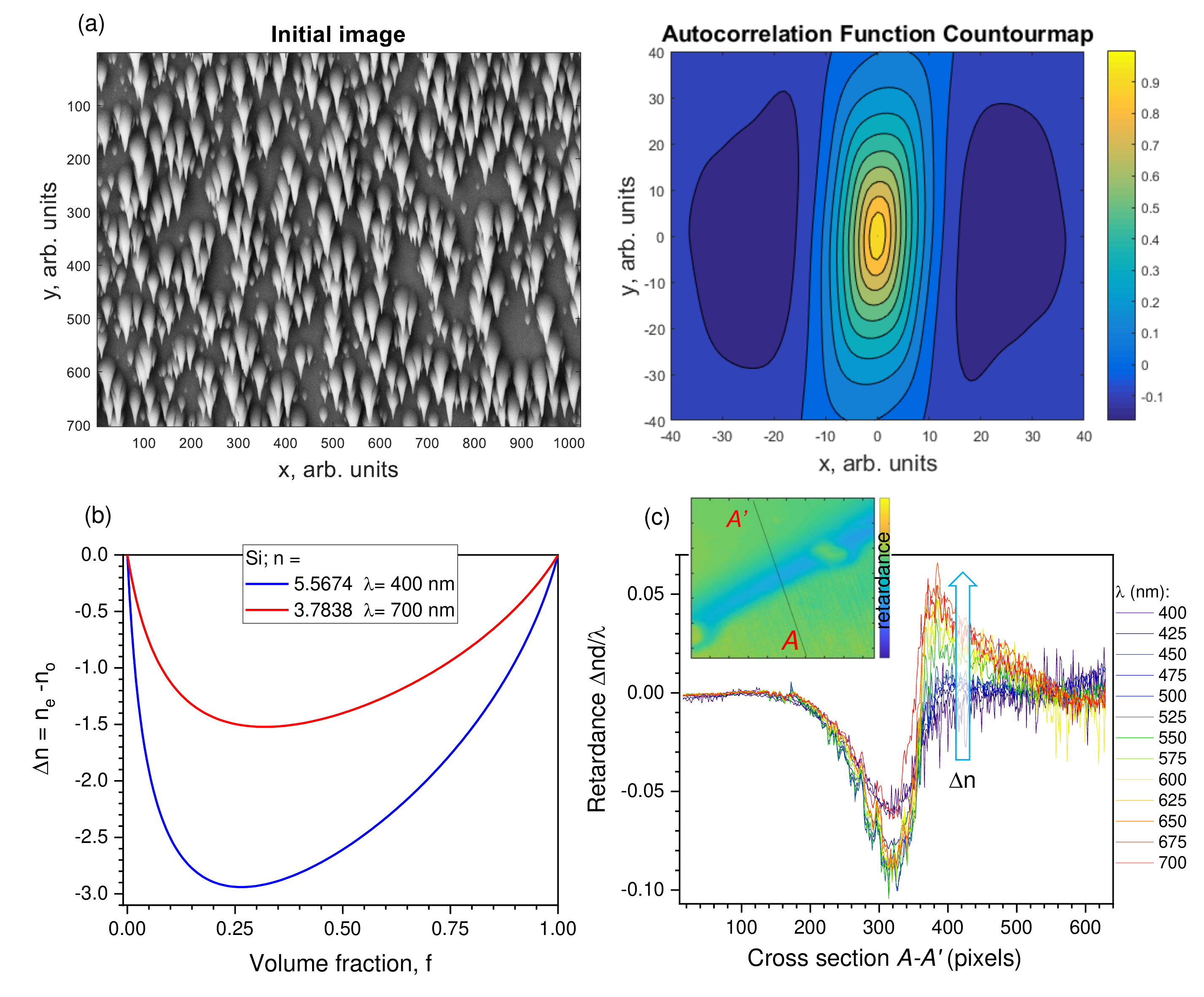}
\caption{(a) SEM image and its auto-correlation function for determining the filing $f$ and the anisotropy $k$ factors. Slight asymmetry of the auto-correlation plot is due to contrast gradient in the  SEM image. (b) The form birefringence $\Delta n = n_e - n_o$ of air-Si with changing volume ratio, $f$, at two wavelengths used in experiments calculated for the grating structure Eqn.~\ref{En}; the refractive index of Si was taken from database ~\cite{refind}. See Appendix for estimates by the Bruggeman EMT (Eqn.~\ref{Brugeman}). The form birefringence is negative due to $n_e<n_o$. (c) Retardance taken with 13 filters with 20~nm bandwidth along the same cross section line; color pallet corresponds to the filter color. The region where from birefringence dominated retardance changes is shown by arrow (see Figs.~\ref{f-reta}(a,b) and \ref{f-sect}). The inset shows retardance map of b-Si at 425~nm wavelength.} \label{f-fobi}
\end{figure}

\clearpage 
\bibliography{paper6c}

\end{document}